# Zeeman Spin-Splitting in the (010) β-Ga$_2$O$_3$ Two-Dimensional Electron Gas

Adam T. Neal[1,*], Yuewei Zhang[2,†], Said Elhamri[1,3], Siddharth Rajan[2], Shin Mou[1,*]

[1] Air Force Research Laboratory, Materials and Manufacturing Directorate, Wright Patterson AFB, OH, USA 45433
[2] The Ohio State University, Department of Electrical and Computer Engineering, Columbus, OH, USA 43210
[3] University of Dayton, Department of Physics, Dayton, OH, USA 45469

[†] Present Address: University of California Santa Barbara, Santa Barbara, CA, USA 93106
[*] Electronic Address: adam.neal.3@us.af.mil, shin.mou.1@us.af.mil

Abstract:

Through magneto-transport measurements and analysis of the observed Shubnikov de Haas oscillations in (010) (Al$_x$Ga$_{1-x}$)$_2$O$_3$/Ga$_2$O$_3$ heterostructures, spin-splitting of the Landau levels in the (010) Ga$_2$O$_3$ two-dimensional electron gas (2DEG) has been studied. Analysis indicates that the spin-splitting results from the Zeeman effect. By fitting the both the first and second harmonic of the oscillations as a function of magnetic field, we determine the magnitude of the Zeeman splitting to be 0.4 $\hbar\omega_c$, with a corresponding effective g-factor of 2.7, for magnetic field perpendicular to the 2DEG.

Since the first proof-of-concept device demonstrations, [1,2] interest in ultra-wide bandgap $Ga_2O_3$ as a transistor material for power electronics has surged due to its large breakdown electric field, experimentally demonstrated to be larger than that of GaN and SiC. [3] While theoretically estimated to be as high as 8 MV/cm, [1] the best experimentally demonstrated peak breakdown field is estimated to be about 6 MV/cm so far. [4] While this high breakdown field will certainly be explored for high voltage vertical device applications, lateral $Ga_2O_3$ devices can also take advantage of this large breakdown electric field through device scaling for applications such as integrated power electronics and radio-frequency electronics. [5,6,7] Motivated by the success of the arsenide MODFETs and nitride based HEMTs, modulation doping of $(Al_xGa_{1-x})_2O_3/Ga_2O_3$ heterostructures has been developed in order to simultaneously achieve maximized mobility and high carrier density in lateral devices. [8,9] The community, however, is just beginning to understand the fundamental transport properties of the $Ga_2O_3$ two-dimensional electron gas (2DEG) which is the basis of these MODFETs. Initial work analyzed the temperature dependent low-field mobility and Shubnikov-de Haas oscillations, experimentally determining the cyclotron effective mass, transport, and quantum scattering times. [9] While spin-splitting of the oscillations was observed, it was not analyzed in detail. Namely, the mechanism of the spin-splitting was not identified nor the magnitude of the spin-splitting analyzed. In this work, we identify the Zeeman effect as the mechanism responsible for spin-splitting in the (010) $Ga_2O_3$ 2DEG through measurement and analysis of the Shubnikov-de Haas oscillations. By fitting the oscillations as a function of magnetic field, we determine the magnitude of the Zeeman splitting to be 0.4 $\hbar\omega_c$, where $\hbar\omega_c$ the Landau level separation and cyclotron orbit energy, with a corresponding effective g-factor of 2.7.

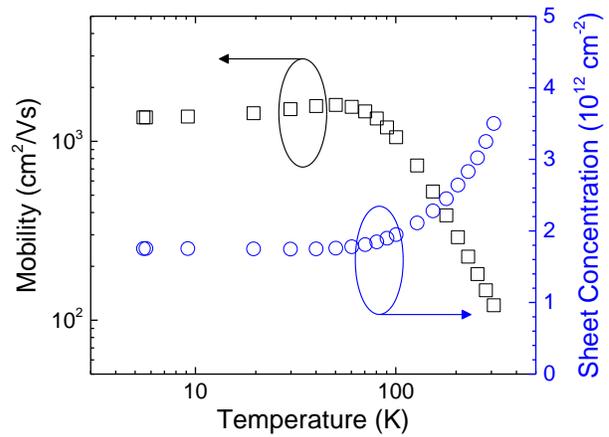

*Figure 1: Temperature dependent mobility and carrier density of Sample 1.*

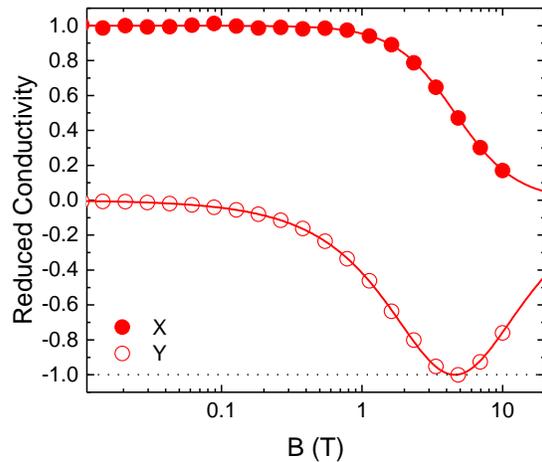

*Figure 2: Magnetic field dependent longitudinal (X) and Hall (Y) normalized conductivity of Sample 1 at T = 30 K. The fit indicates the presence of a single carrier upon cooling.*

We begin by presenting temperature and magnetic field dependent transport data to outline the basic properties of the two samples characterized in this work. Both Sample 1 and Sample 2 were microfabricated van der Pauw structures circular in shape, and details of the heterostructure growth and sample fabrication are published elsewhere. [9] Figure 1 shows temperature dependent carrier density and mobility for Sample 1, determined to be $1.75\times10^{12}$ cm$^{-3}$ and 1360 cm$^2$/Vs, respectively, at T = 5.5 K. The elevated carrier density at higher temperatures indicates the presence of parallel conduction in a parasitic channel, likely the Si modulation dopants in the AlGaO barrier. While this parasitic conduction makes this particular sample unsuitable for transistor fabrication, this parallel channel freezes out below about T = 100 K as shown in Figure 1, allowing us to characterize the intrinsic properties of the 2DEG at low temperatures. Multicarrier analysis of the magnetic field dependent longitudinal and Hall conductivity at T = 30 K, shown in Figure 2, indicates the presence of a single charge carrier, consistent with the freeze out of the parasitic channel with reduced temperature. Temperature dependent transport data for Sample 2 (not shown) has been previously reported, with a carrier density of $2.1 \times 10^{12}$ cm$^{-3}$ and peak mobility of 2800 cm$^2$/Vs at T = 50 K, with no evidence of parallel conduction. [9]

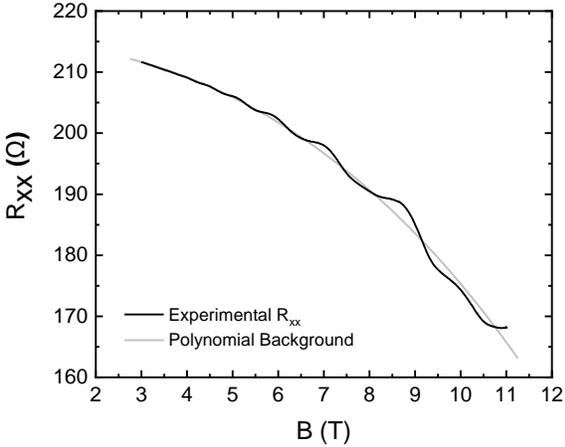

*Figure 3: Magnetic field dependent longitudinal resistance of Sample 1 at T = 1.5K, showing SdH oscillations.*

Figure 3 shows the longitudinal resistance as a function of magnetic field for Sample 1 at a temperature of 1.5 K, with oscillations clearly visible above 5 Tesla. To better highlight the oscillations, a polynomial background, also shown in Figure 3, was subtracted from the experimental data, with the result plotted as a function of reciprocal magnetic field in Figure 4 for Sample 1, a new experimental measurement for this work. Similar data for Sample 2 is shown in Figure 5. Note that the experimental data for Sample 2 was also part of our previous publication. [9] The oscillation periodicity in reciprocal magnetic field, characteristic of Shubnikov-de Haas oscillations and Landau level formation, is clearly observed. There is a clear similarity in the oscillations observed in both samples, aside from the different oscillations period resulting from their different carrier densities. We note that the experimental data for Sample 1 and that for Sample 2 were measured in two different laboratories, confirming the repeatability of the results. Most notably, as magnetic field increases (reciprocal magnetic field decreases), new maxima in the oscillations emerge around $B^{-1}$ ~ 0.10 T$^{-1}$ for Sample 1 and $B^{-1}$ ~ 0.09 T$^{-1}$ for Sample 2, superimposed over an SdH minima also at those positions. As we will show, this emerging maxima corresponds to the second harmonic of the Shubnikov-de Haas oscillations, appearing due to the suppression of the SdH first harmonic due to Zeeman spin-splitting.

In order to establish the Zeeman nature of the spin-splitting in the (010) Ga$_2$O$_3$ 2DEG, it is useful to introduce a generalized model which describes the amplitude and shape of the SdH oscillations as a function of magnetic field and spin-splitting. While quantitative models describing the SdH oscillations up to the first harmonic have existed for some time, [10] a modern formulation by Tarasenko [11,12] extended this quantitative description up to the second harmonic which will be useful here. Following Tarasenko, [11,12] the magnitude of the first and second harmonics of the SdH oscillations are proportional to:

$$\delta_1 = 2\exp\left(-\frac{\pi}{\omega_c\tau}\right)\cos\left(2\pi\frac{E_F}{\hbar\omega_c} - \pi + \phi_1\right)\cos\left(\pi\frac{\Delta}{\hbar\omega_c}\right) \quad (1)$$

$$\delta_2 = 2\exp\left(-\frac{2\pi}{\omega_c\tau}\right)\cos\left(4\pi\frac{E_F}{\hbar\omega_c} - 2\pi + 2\phi_2\right)\cos\left(2\pi\frac{\Delta}{\hbar\omega_c}\right) \quad (2)$$

where $\delta_1$ refers to the first harmonic, $\delta_2$ the second harmonic, $\hbar\omega_c$ the Landau level energy separation or cyclotron orbit energy, $E_F$ the fermi level, $\tau$ the quantum scattering time, $\phi_1$ and $\phi_2$ phases of the SdH oscillations, and $\Delta$ the spin-splitting energy. Note that this model is very general, describing SdH oscillations subject to Zeeman splitting, spin-orbit splitting, or some combination of these effects by assuming the appropriate magnetic field dependence for $\Delta$.

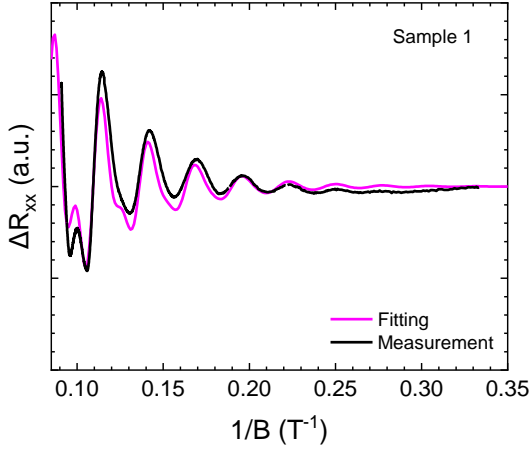

*Figure 4: SdH oscillations of Sample 1 plotted versus reciprocal magnetic field, with a polynomial background subtracted.*

With the aid of this model, let us consider some qualitative features of these Shubnikov-de Haas oscillations to determine the nature of the spin-splitting. First, examining the oscillations shown in Figure 4 and Figure 5, there is a clear lack of two-frequency beating, ruling out spin-orbit coupling as the origin of the spin-splitting. For spin-orbit coupling induced splitting, $\Delta$ is independent of magnetic field, meaning that multiplication by the term $\cos\left(\pi\frac{\Delta}{\hbar\omega_c}\right)$ produces oscillations at two frequencies, $\frac{E_F \pm \Delta/2}{\hbar\omega_c/B}$, which correspond to differing areas of the spin-split Fermi-surfaces produced by spin-orbit coupling. These two oscillation frequencies produce beating effects in the SdH oscillations, which are not observed in Figure 4 and Figure 5. By contrast, Zeeman splitting is proportional to magnetic field, just like the Landau level splitting $\hbar\omega_c$, so the term $\cos\left(\pi\frac{\Delta}{\hbar\omega_c}\right)$ simply leads to a phase difference of $\frac{\Delta}{\hbar\omega_c}$ for Zeeman split Landau levels, consistent with the lack of beating in the SdH oscillations observed here in the (010) $Ga_2O_3$ 2DEG. Second, the case for Zeeman splitting is further strengthened when considering the consistent shape of the SdH oscillations as a function of reciprocal magnetic field from one period to the next. Most clearly seen in the two minima at $B^{-1} \sim 0.09$ T$^{-1}$ and $B^{-1} \sim 0.117$ T$^{-1}$ in Figure 5, the position of the emerging maxima is consistent from one period to the next, becoming better resolved as the increasing magnetic field strengthens localization in the system and enhances the spin-splitting. As previously mentioned, the Landau level separation $\hbar\omega_c$ and Zeeman induced spin-splitting $\Delta$ are both proportional to magnetic field, meaning that their ratio, $\frac{\Delta}{\hbar\omega_c}$, remains constant as magnetic field increases, resulting in the consistent SdH oscillation shape from one period to the next. Finally, as the data in our previous publication shows, [9] there is a clear modulation of the spin-splitting as a function of the angle of the magnetic field, also consistent with Zeeman splitting. This modulation with angle occurs because the SdH Landau level separation $\hbar\omega_c$ depends on the normal component of the magnetic field while the Zeeman splitting $\Delta$ depends on the total magnetic field. By changing the ratio $\frac{\Delta}{\hbar\omega_c}$ with the angle, the relative magnitude of the first and second harmonics changes, changing the shape of the SdH oscillations. With these three qualitative features observed in the SdH oscillations, we can confidently conclude that the observed spin-splitting in the (010) Ga$_2$O$_3$ 2DEG is the result of Zeeman effect.

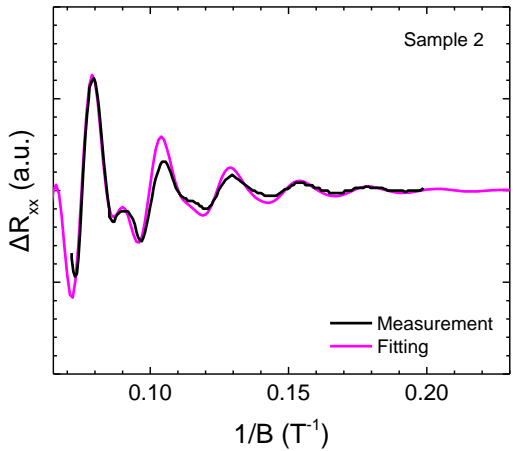

*Figure 5: SdH oscillations of Sample 2 plotted versus reciprocal magnetic field, with a polynomial background subtracted.*

To quantitatively determine the magnitude of the Zeeman spin-splitting, a few approaches have been adopted previously in the literature. [10,13] In high mobility samples with minimal broadening of the Landau levels and robust localization such that the longitudinal magnetoresistance approaches zero between the Landau levels, the splitting of the Landau level manifests very obviously as the splitting of the SdH maxima. In this case, the spin-splitting energy can be calculated directly from the separation of the peaks in reciprocal magnetic field. However, the situation is different for our (010) Ga$_2$O$_3$ 2DEG studied here. Given the relatively early stage of development for the growth of the AlGaO/GaO heterostructure, the relatively modest low temperature mobility of the samples leads to significant broadening of the Landau levels, broadening larger than the Landau level separation. In this case, rather than sharp peaks, the SdH effect manifests as oscillations, and it is no longer possible to resolve individual spin-split Landau levels due to the broadening. For these weaker oscillations, a tilted magnetic field is often used to extract the magnitude of the spin-splitting by changing the relative spin-splitting, $\frac{\Delta}{\hbar\omega_c}$, as discussed previously. One can then determine the angle at which the first harmonic of the SdH oscillation changes phase or, if resolved, the second harmonic dominates the SdH oscillations. At this angle, the spin-splitting and the Landau level separation are exactly equal, enabling one to back-calculate the spin-splitting for perpendicular magnetic field. [13] However, isotropy of the g-factor is assumed in such an analysis, which may not be appropriate for a 2DEG with strong confinement in one spatial direction or a highly anisotropic crystal like Ga$_2$O$_3$.

Therefore, we have adopted an approach in which fitting of the oscillations as a function of magnetic field can be used to extract the spin-splitting energy [10] using the previously introduced formulation of Tarasenko. [11,12] In this approach, fitting of the SdH oscillations as a function of magnetic field allows one to determine the relative amplitudes of the first and second harmonic, which determines the ratio $\frac{\Delta}{\hbar\omega_c}$. Other fitting parameters in the model include the oscillation frequency $\frac{E_F}{\hbar\omega_c/B}$, quantum scattering time $\tau$, and phase factors $\phi_1$ and $\phi_2$. The quantum scattering time, $\tau$, was determined from Dingle plots of the oscillation amplitude, quantified as the differences between adjacent minima and maxima, as a function of magnetic field prior to fitting the other parameters. The resulting fits are pictured in Figure 4 and Figure 5, with the fitting parameters summarized in Table 1. The indicated errors were calculated based on a 30% increase in the sum of the squared residuals. Based on the fitting, we find that the magnitude of the Zeeman splitting, normalized to the Landau level separation, is approximately 0.4 for both Sample 1 and Sample 2 at normal incidence of the external magnetic field. Furthermore, based on this spin-splitting, the effective g-factor for Zeeman splitting for magnetic field perpendicular to the (010) 2DEG can be calculated as $g = 2\left(\frac{\Delta}{\hbar\omega_c}\right)\left(\frac{m_o}{m_*}\right)$, which is approximately 2.7 assuming an effective mass of 0.3 $m_o$. [9] Electron paramagnetic resonance (EPR) measurements have also been used to study the g-factor of electronic states in gallium oxide; however, there is some debate in the literature as to the origin of an EPR signal attributed either to delocalized conduction band electrons or to shallow donor states. [14,15] In any case, estimates of the g-factor for this state from EPR measurements yield a value of approximately 2, not too different from the free electron. We attribute our observation of a g-factor higher than 2 to enhancement of the effective g-factor by exchange interaction. [16,17,18] We do note the unusual phases for the first harmonic, $\phi_1$, and second harmonic, $\phi_2$, which consistently differ from the expected value of zero. While this topic is beyond the scope of this letter, it is an interesting topic for future investigation.

Table 1: SdH fitting parameters

|  | | Sample 1 | Sample 2 |
|---|---|---|---|
| $\frac{\Delta}{\hbar\omega_c}$ | Tesla | 0.40±0.01 | 0.43±0.01 |
| $\frac{E_{FB}}{\hbar\omega_c}$ | Tesla | 36.9±0.1 | 40.4±0.1 |
| $\tau$ | $10^{-13}$ s | 2.1±0.4 | 1.9±0.4 |
| $\phi_1$ | radians | $\pi/2\pm\pi/32$ | $\pi/2\pm\pi/32$ |
| $\phi_2$ | radians | $5\pi/8\pm\pi/32$ | $5\pi/8\pm\pi/32$ |

In conclusion, we have characterized the spin-splitting of Landau levels in in the (010) Ga2O3 2DEG, establishing the Zeeman nature of the splitting. Through fitting of the first and second harmonics of the SdH oscillations, the Zeeman splitting is determined to be 0.4 ℏω_c, with an effective g-factor of 2.7, for magnetic field perpendicular to the 2DEG.

Acknowledgements:

This material is partially based upon the work supported by the Air Force Office of Scientific Research under Award No. FA9550-18RYCOR098. S.R. acknowledges support from Air Force Office of Scientific Research under Award No. FA9550-18-1-0479 (GAME MURI). The content of the information does not necessarily reflect the position or the policy of the federal government, and no official endorsement should be inferred.